\documentclass[prc,aps,amsmath,amssymb,nofootinbib,superscriptaddress,12pt]{revtex4-2}
\usepackage{epsfig}
\usepackage[bookmarksnumbered,bookmarksopen,colorlinks,citecolor=blue,linkcolor=blue]{hyperref}
\usepackage{amsmath}
\usepackage{amssymb}
\usepackage{tipa}

\begin{document}

    \title{Refinement of an analytical capture cross section formula }

\author{Ning Wang}
\email{wangning@gxnu.edu.cn}\affiliation{ Guangxi Normal University, Guilin 541004, People's Republic of
	China }
\affiliation{ Guangxi Key Laboratory of Nuclear Physics and Technology, Guilin 541004, People's Republic of
	China }

    \begin{abstract}
        An analytical formula with high accuracy is proposed for a systematic description of the capture cross sections at near-barrier energies from light to superheavy reaction systems. Based on the empirical barrier distribution (EBD) method, three key input quantities are refined by introducing nuclear surface correction to the Coulomb parameter $z$ for calculating the barrier height, incorporating the reaction $Q$-value and shell correction into the barrier distribution width calculations, and considering the deep inelastic scattering effects of superheavy systems on the barrier radius. With these refinements, not only the accuracy of the calculated barrier height but also the accuracy of the predicted capture cross sections, is substantially improved. The average deviation (in logarithmic scale) between the predicted cross sections and the experimental data for a total of 426 reaction systems with $ 35 < Z_1 Z_2 < 2600$ is sharply reduced from 3.485 to 0.113. 
      
    \end{abstract}

     \maketitle

\newpage
  
   \begin{center}
  	\textbf{ I. INTRODUCTION }\\
  \end{center}

  The study of heavy-ion fusion reactions at near-barrier energies \cite{Wong73,Bass74,Swiat82,BW91,Gup92,Lei95,Mort99} is of significant importance in nuclear physics, particularly for the synthesis of superheavy nuclei (SHN) \cite{Hof00,Hof04,Mori04a,Ogan10,Ogan15,Mori20,Adam04} and the exploration of nuclear structure effects \cite{Stok78,Das98,Wolski,Sarg11,Jia14}. A critical aspect of the study is the accurate calculation of the capture cross sections, which are influenced by complex nuclear structure effects and dynamical processes such as dynamical deformation and nucleon transfer. These factors couple with the relative motion of the colliding nuclei \cite{Hag99,Dasso87}, necessitating sophisticated models to describe the fusion dynamics accurately.
  
 One of the widely used approaches \cite{Zag01,SW04,liumin,Wang09,Wangbing,Jiang22} to account for these couplings is the introduction of an empirical barrier distribution (EBD). By using a single Gaussian function to parameterize the barrier distribution, the capture cross sections can be expressed as an analytical formula \cite{SW04,Cap11}, which is named as EBD method in code KEWPIE2 \cite{EBD}. It is found that the EBD method has demonstrated reasonable success for heavy fusion systems  \cite{EBD,EP} and provides a robust framework for analyzing cold fusion leading to the synthesis of superheavy nuclei \cite{Cap11}. While the EBD method works well for heavy systems, its accuracy diminishes for fusion reactions with lighter nuclei due to oversimplifications. Firstly, the parameters in the original EBD method does not explicitly consider the influence of the surface effects of light nuclei on the barrier height and radius, which leads to an over-prediction of the barrier height and under-prediction of the barrier radius for light fusion system such as $^{16}$O+$^{16}$O. Furthermore, the model neglects the influence of the reaction $Q$-value on the cross section. For instance, the fusion systems $^{132}$Sn+$^{40,48}$Ca, despite both involving spherical nuclei, exhibit significant  difference of the sub-barrier cross sections due to their distinct reaction $Q$-values. The influence of the $Q$-value on the width of the barrier distribution, which is clearly observed in \cite{Yao24}, is not involved in the EBD method since the width parameter is only dependent on the deformations of the reaction partners and the average barrier height. Neglecting $Q$-value effects in the EBD calculations could result in that the data of both reactions $^{132}$Sn+$^{40,48}$Ca cannot be well reproduced simultaneously. In addition, the influence of deep inelastic scattering (DIS) on the barrier radius for the superheavy systems  observed in \cite{Wang25,Koz16}, is also neglected in the EBD method.  
  
 To address its limitations for light systems, we present an enhanced version of the EBD formula. The improvements include incorporating the surface effects of light nuclei and the contribution of the competition between shell effect and isospin effect to the barrier height. Additionally, we consider the influence of the reaction $Q$-value on the barrier distribution width and account for deep inelastic scattering effects on the barrier radius for superheavy systems. These refinements substantially enhance the model's accuracy, enabling a more comprehensive description of fusion reactions across a wide range of systems, from light to superheavy cases.
  
 The structure of this paper is as follows: In Sec. II, the framework of EBD method and the three key input quantities will be introduced. In Sec. III, the results from the proposed formula for a series of reaction systems and some discussions are presented.  Finally a summary is given in Sec. IV.

 \begin{center}
	\textbf{ II. EMPIRICAL BARRIER DISTRIBUTION FORMULA }\\
\end{center}
  
 In the EBD method, the capture cross section is written as \cite{SW04,Cap11,EBD},
 \begin{equation}
 	\sigma_{\rm {cap} } (E_{\rm c.m.})=\pi R_B^{2}  \frac{W}{\sqrt{2}E_{\rm c.m.}}
 	[X  {\rm erfc}(-X)+\frac{1}{\sqrt{\pi}}\exp(-X^{2}) ],
 \end{equation}
 where $X = (E_{\rm c.m.}-V_B)/\sqrt{2}W $.  $E_{\rm c.m.}$ denotes the center-of-mass incident energy. $V_B$ and $W$ denote the centroid and the standard deviation of the Gaussian function, respectively.  $R_B$ denotes the barrier radius.
 
 In this work, a new version EBD2 of the formula is proposed by refining the three input quantities: $V_B$, $R_B$ and $W$. In Ref. \cite{Wen22}, Wen \textit{et al} found that the extracted barrier heights $V_B$  are approximately linear proportional to the Coulomb parameter $Z_1 Z_2/(A_1^{1/3}+A_2^{1/3})$. In EBD, $V_B$ is parameterized by a cubic polynomial of the Coulomb parameter \cite{Cap11}. In this work, the average barrier height $V_B$ (in MeV) is parameterized as, 
   \begin{eqnarray}
 	V_B=1.051 z +0.000335 z^2 + \Delta B
 \end{eqnarray}
with the Coulomb parameter re-written as,
  \begin{eqnarray}
 	z=\frac{Z_1 Z_2}{A_1^{1/3}+A_2^{1/3}} F_S
 \end{eqnarray} 
Here, $Z_1$ and $Z_2$  denote the charge number of the projectile nucleus and that of the target, respectively. $A_1$ and $A_2$  denote the corresponding mass number of the reaction partners. The surface correction factor $F_S=1-Z_1^{-1/3} Z_2^{-1/3}$ is introduced to consider the surface effects of light nuclei and the Coulomb exchange term \cite{Wang10,Wan25}. The correction term $\Delta B=\sum {\Delta_i I_i^2}$ in Eq.(2) is to consider the competition between shell effect and isospin effect, with the shell gap  $\Delta_i$ \cite{Mo16} and isospin asymmetry $I_i=(N_i-Z_i)/A_i$ of the reaction partners ($i=1$ for projectile and $i=2$ for target). It is known that on one hand, the neutron transfer and neck formation in reaction process with neutron-rich nuclei can lower the capture barrier and thus enhance the fusion cross sections at sub-barrier energies \cite{Stelson88,liumin}. On the other hand, for fusion reactions between magic nuclei, the strong shell effect inhibits the lowering barrier effect \cite{liumin}. This competition is evident for the reactions with neutron-rich nuclei $^{48}$Ca and $^{132}$Sn.

 The average barrier radius is written as
  \begin{eqnarray}
R_B=r_0 (A_1^{1/3}+A_2^{1/3})  (F_S F_{\rm DIS})^{-1},
\end{eqnarray} 
with $r_0=1.10 $ fm. The factor $F_{\rm DIS}= 1+\exp(X_{\rm B}-X_0) $ with the Bass parameter \cite{Bass74} $X_{\rm B}=\frac{Z_1 Z_2}{(A_1^{1/3}+A_2^{1/3})A_1^{1/3}A_2^{1/3}}$ is introduced to consider the influence of deep inelastic scattering on the barrier radius in superheavy systems inspired by the results in \cite{Wang25,Koz16}. The dimensionless parameter $X_0=10.2$ is determined by the measured capture cross sections of some super-heavy systems \cite{Itkis22} such as $^{64}$Ni + $^{238}$U. Bass concluded that fusion is excluded for the systems with $X_{\rm B}>12.6$ \cite{Bass74} due to disappearance of the capture pocket.

The standard deviation of the Gaussian function is parameterized as, 
  \begin{eqnarray}
	W=c_0 (1+w_{d}) + c_1 V_B  \sqrt{w_1^2+w_2^2+w_0^2} -\Delta B/N_{\rm CN}^{1/3},
\end{eqnarray} 
where $w_{d}=\sum {|\beta_{2i}| A_i^{1/3}}$ and $w_i = r_0 A_i^{1/3} \beta_{2i}^2/4\pi$ \cite{Cap11,EBD}, with the mass number $A_1$ and $A_2$ of the reaction partners, and their quadrupole deformation parameters $\beta_2$ taken from the WS4 model \cite{WS4} for prolate nuclei heavier than $^{16}$O. For oblate nuclei ($\beta_2<0$) or those with hexadecapole deformation $\beta_4<0$, the corresponding deformation parameter is set as $\sqrt{\beta_2^2+\beta_4^2}/2$.  In this work, the $Q$-value dependence of the parameter $w_0=(V_B+Q)/c_2$ is introduced to consider the dynamical effects due to the excitation energy of the reaction system at the capture position which is approximately proportional to the excitation energy of the compound nuclei. $N_{\rm CN}$ denotes the neutron number of the compound nucleus. The optimal values for the four parameters $c_0=0.63$ MeV, $c_1=0.015$ fm$^{-1}$, $c_2=33.0$ MeV$\cdot$fm$^{-1}$ and  $r_0=1.10$ fm in Eq.(5) are obtained by varying these parameters and searching for the minimal deviation between the predicted cross sections and the experimental data for 426 fusion reactions (including the systems collected in Refs. \cite{Wangbing,Itkis22,Chen23} except those induced by nuclei lighter than $^{12}$C).  For $\alpha$-induced fusion reactions, the structure effects could be neglected and the value of $W$ is simply written as $W=c_0 + c_1 V_B$.

By inputting the values of $V_B$, $R_B$ and $W$ obtained from Eqs.(2)-(5) under the fixed parameters into Eq.(1), the capture cross sections of any selected heavy-ion fusion reactions at energies around the Coulomb barrier can be directly calculated based on the EBD2 formula.

\begin{center}
	\textbf{ III. RESULTS AND DISCUSSIONS }\\
\end{center}

\begin{figure}
	\setlength{\abovecaptionskip}{ 0.  cm}
	\includegraphics[angle=0,width=0.7 \textwidth]{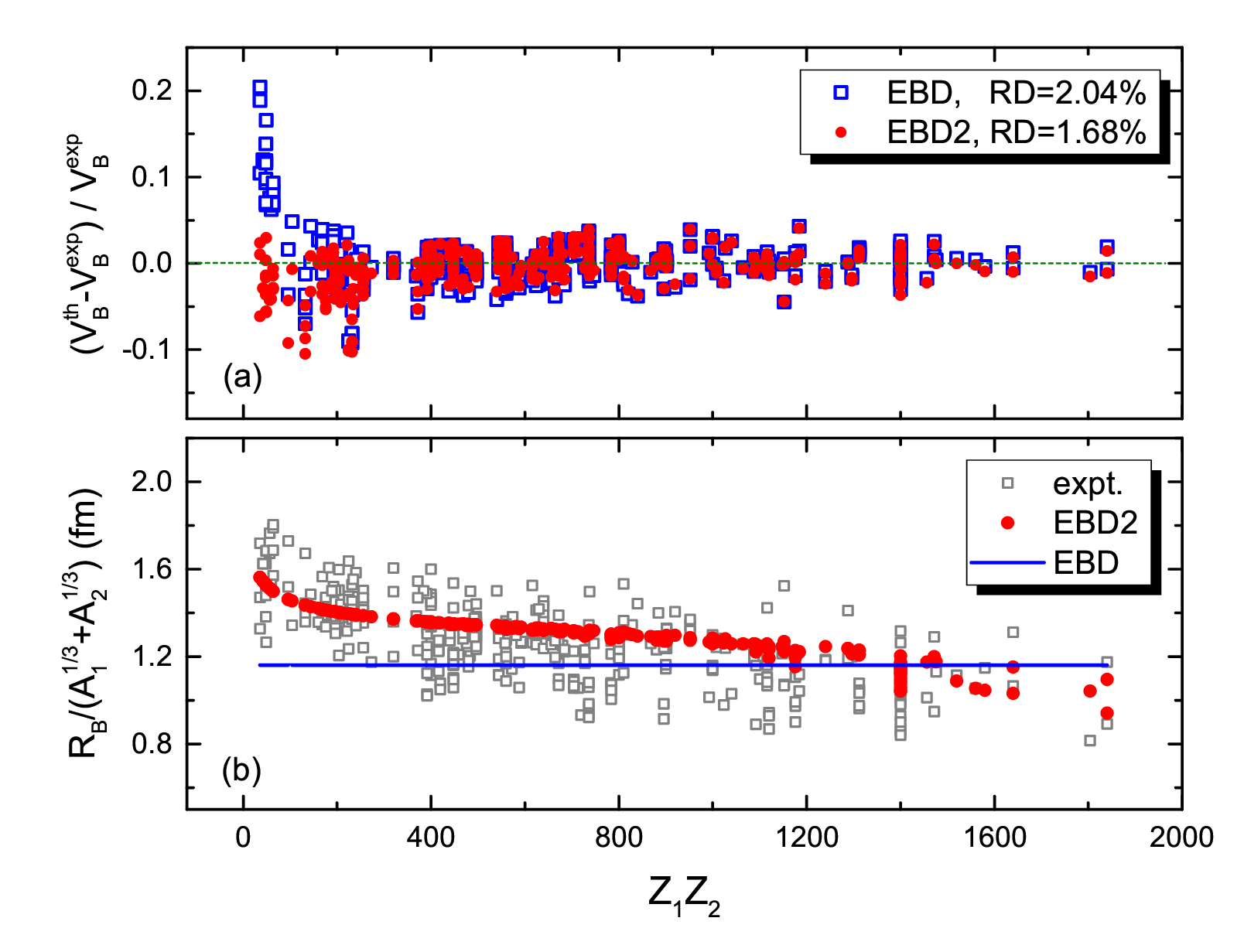}
	\caption{ (a) Relative deviations between the calculated barrier heights $V_B^{\rm th}$ and 382 extracted barrier heights  \cite{Chen23}  $V_B^{\rm exp}$ for reactions induced by nuclei with $Z_1\ge 6$ and $Z_2\ge 6$. The squares and the circles denote the results of EBD \cite{Cap11,EBD} and those of EBD2, respectively.  (b) Barrier radii $R_B$ scaled by $A_1^{1/3}+A_2^{1/3}$. The squares and the circles denote the extracted reduced barrier radii \cite{Chen23} and the predictions of EBD2 by Eq.(4), respectively.  The line denotes the results of EBD \cite{Cap11,EBD}.
	}
\end{figure}

     \begin{table}    	
    	
 	\caption{ Barrier parameters according to Eqs.(2)-(5) for 20 fusion reactions. $\chi_{\log}^{2}$ denotes the deviation between measured cross sections and the predicted $\sigma_{\rm {cap} }$ with EBD2. }
 	\begin{tabular}{ccccc}
 		\hline\hline
 		
 		~~~Reaction~~~  & ~~~$V_B$ (MeV)~~~ &~~~$R_B$ (fm)~~~  &  ~~~W (MeV)~~~  & ~~~$\chi_{\log}^{2}$~~~  \\
 		\hline
 		$^{16}$O+$^{16}$O    &  10.04   &  7.39   &  0.75   &  0.041  	\\
 		$^{18}$O+$^{58}$Ni   &  30.69   &  8.54   &  1.44   &  0.011  	\\
 		$^{16}$O+$^{92}$Zr   &  41.39   &  9.04   &  1.36   &  0.039  	\\
 		$^{16}$O+$^{154}$Sm  &  58.93   &  9.88   &  2.58   &  0.007  	\\
 		$^{16}$O+$^{208}$Pb  &  74.35   &  10.43  &  1.48   &  0.109	\\
 		$^{16}$O+$^{238}$U   &  80.92   &  10.69  &  3.03   &  0.008  	\\
 		 		
 		$^{40}$Ca+$^{48}$Ca  &  52.67   &  8.95   &  1.94   &  0.023     \\
 		$^{40}$Ca+$^{132}$Sn &  115.79  &  10.08  &  3.83   &  0.030     \\
 		
 		$^{48}$Ca+$^{132}$Sn &  113.24  &  10.47  &  2.33   &  0.014     \\
 		$^{48}$Ca+$^{154}$Sm &  137.12  &  10.45  &  4.29   &  0.011     \\
 		$^{48}$Ca+$^{208}$Pb &  174.16  &  10.38  &  2.10   &  0.185     \\
 		$^{48}$Ca+$^{238}$U  &  191.03  &  10.23  &  4.07   &  0.012     \\
 		$^{48}$Ca+$^{248}$Cm &  198.22  &  10.06  &  4.09   &  0.094     \\
 		 		 		 		 		 
 		$^{54}$Cr+$^{208}$Pb &  208.39  &  8.80   &  2.84   &  $-$        \\
		$^{54}$Cr+$^{238}$U  &  228.93  &  8.08   &  5.24   &  $-$        \\ 		
		$^{54}$Cr+$^{244}$Pu &  233.07  &  7.91   &  5.26   &  $-$        \\ 	
		$^{54}$Cr+$^{243}$Am &  235.98  &  7.57   &  4.92   &  $-$        \\ 	
		$^{54}$Cr+$^{248}$Cm &  237.65  &  7.62   &  5.01   &  $-$        \\ 	
		
		$^{64}$Ni+$^{208}$Pb &  240.66  &  7.18   &  2.42   &  $-$        \\ 	
		$^{64}$Ni+$^{238}$U  &  264.75  &  6.08   &  4.76   &  0.136      \\ 	
								 		
 		\hline\hline
 	\end{tabular}
 \end{table}

To quantify the impact of the proposed refinements, we first evaluate the accuracy of the two-parameter barrier height formula, i.e., Eq.(2). Fig. 1(a) shows the relative deviations between the calculated barrier heights $V_B^{\rm th}$ according to Eq.(2) and 382 extracted barrier heights \cite{Chen23} $V_B^{\rm exp}$ from the measured fusion excitation functions. One sees that the results of EBD2 are much better than that of EBD for light systems ($Z_1 Z_2\le 200$). The mean value of the relative deviations (RD)  in absolute value with respect to the 382 barrier heights is only $1.65\%$ with EBD2, which is much smaller than that of the three-parameter formula in EBD (${\rm RD} =2.04\%$) and that of two-parameter MCW potential \cite{Wen22} (${\rm RD} =1.95\%$). We note that the accuracy of the barrier height formula can be evidently improved in EBD2 by introducing the surface correction factor $F_S$. In both EBD and MCW, the Coulomb parameter is expressed as $z=Z_1 Z_2/(A_1^{1/3}+A_2^{1/3})$ neglecting the correction factor $F_S$, and  the deviations between model predictions and data for light fusion system thus significantly increase. Fig. 1(b) shows the barrier radius coefficient, i.e., $R_B/(A_1^{1/3}+A_2^{1/3})$.  The squares denote the extracted data \cite{Chen23} which evidently decrease with the charge product $Z_1 Z_2$. The line and the circles denote the results of EBD (1.16 fm) and those of EBD2, respectively. One can see that the decreasing trend of the data can be remarkably well reproduced by EBD2. In Table I, we list the calculated barrier parameters according to Eqs.(2)-(5) and the deviation between experimental data and the calculated 	$\sigma_{\rm {cap} }$ by using Eq.(1) which will be discussed later. One can see that the barrier radii $R_B$ are systematically reduced for super-heavy systems due to the factor $F_{\rm DIS}$. From the values of $W$, we also note that the barrier distributions are evidently broaden for reactions with well-deformed target nuclei, compared to the reactions induced by spherical ones.

\begin{figure}
	\setlength{\abovecaptionskip}{ 0 cm}
	\includegraphics[angle=0,width=0.9\textwidth]{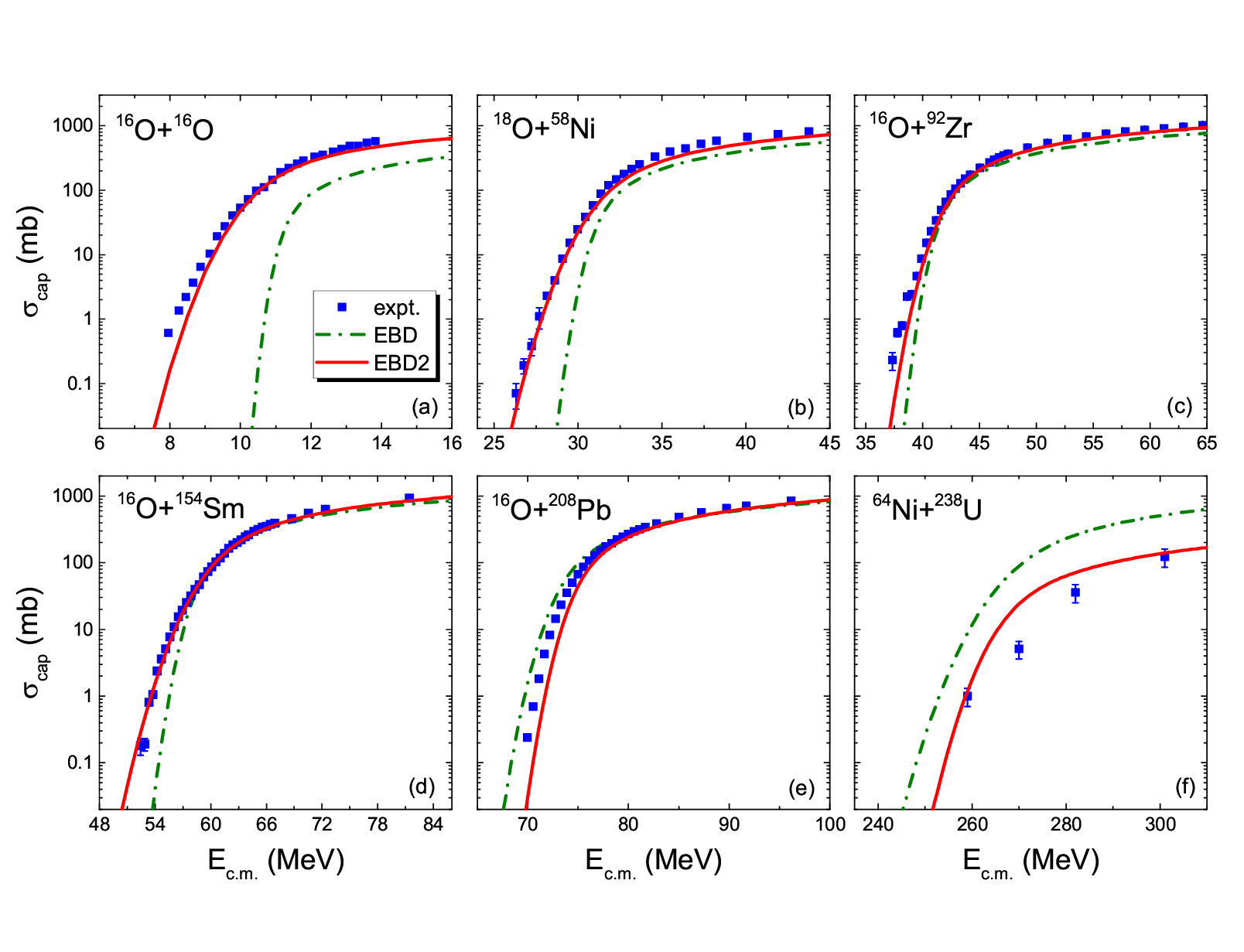}
	\caption{ Comparison of EBD and EBD2 predictions with measured capture cross sections for $^{16}$O+$^{16}$O \cite{Kuro87}, $^{18}$O+$^{58}$Ni \cite{Jia25},  $^{16}$O+$^{92}$Zr \cite{New01},  $^{16}$O+$^{154}$Sm \cite{Lei95},  $^{16}$O+$^{208}$Pb \cite{Mort99}  and $^{64}$Ni+$^{238}$U  \cite{Itkis22}. The squares denote the experimental data. The dot-dashed curves and the solid curves denote the results of EBD and EBD2, respectively. }
\end{figure}

In Fig. 2, we show the predicted capture cross sections for six fusion reactions $^{16}$O+$^{16}$O, $^{18}$O+$^{58}$Ni, $^{16}$O+$^{92}$Zr,  $^{16}$O+$^{154}$Sm,  $^{16}$O+$^{208}$Pb and $^{64}$Ni+$^{238}$U. The dot-dashed curves and the solid curves denote the results of EBD and EBD2, respectively. One sees that the experimental data can be  much better reproduced with EBD2 from light fusion system $^{16}$O+$^{16}$O to superheavy system $^{64}$Ni + $^{238}$U. For $^{16}$O+$^{16}$O, the results of EBD are much smaller than the experimental data by 5 orders of magnitude at $E_{\rm c.m.}=10 $ MeV. The predicted barrier height $V_B=11.02$ MeV with EBD is higher than the extracted one by about $10\%$. Simultanouesly, the predicted barrier radius $R_B=5.85$ fm with EBD is much smaller than the extracted value ($8.52 \pm 0.24$ fm) \cite{Chen23} from the fusion excitation function and the extracted experimental data ($8.63 \pm 0.32$ fm) \cite{Wang18} from the measured elastic scattering angular distribution based on identical-particle interference by about $32\%$ (see Fig. 1(b)), which leads to the significant under-prediction of the fusion cross sections. For the superheavy system $^{64}$Ni+$^{238}$U,  the extracted capture cross sections from the measured mass-total kinetic energy (TKE) distributions are significantly smaller than the predicted results of EBD. EBD2 reproduces the data more accurately since the factor $F_{\rm DIS}$ is introduced in barrier radius to consider the influence of deep inelastic scattering for superheavy systems.

\begin{figure}
	\setlength{\abovecaptionskip}{ -4.5 cm}
	\includegraphics[angle=0,width=1 \textwidth]{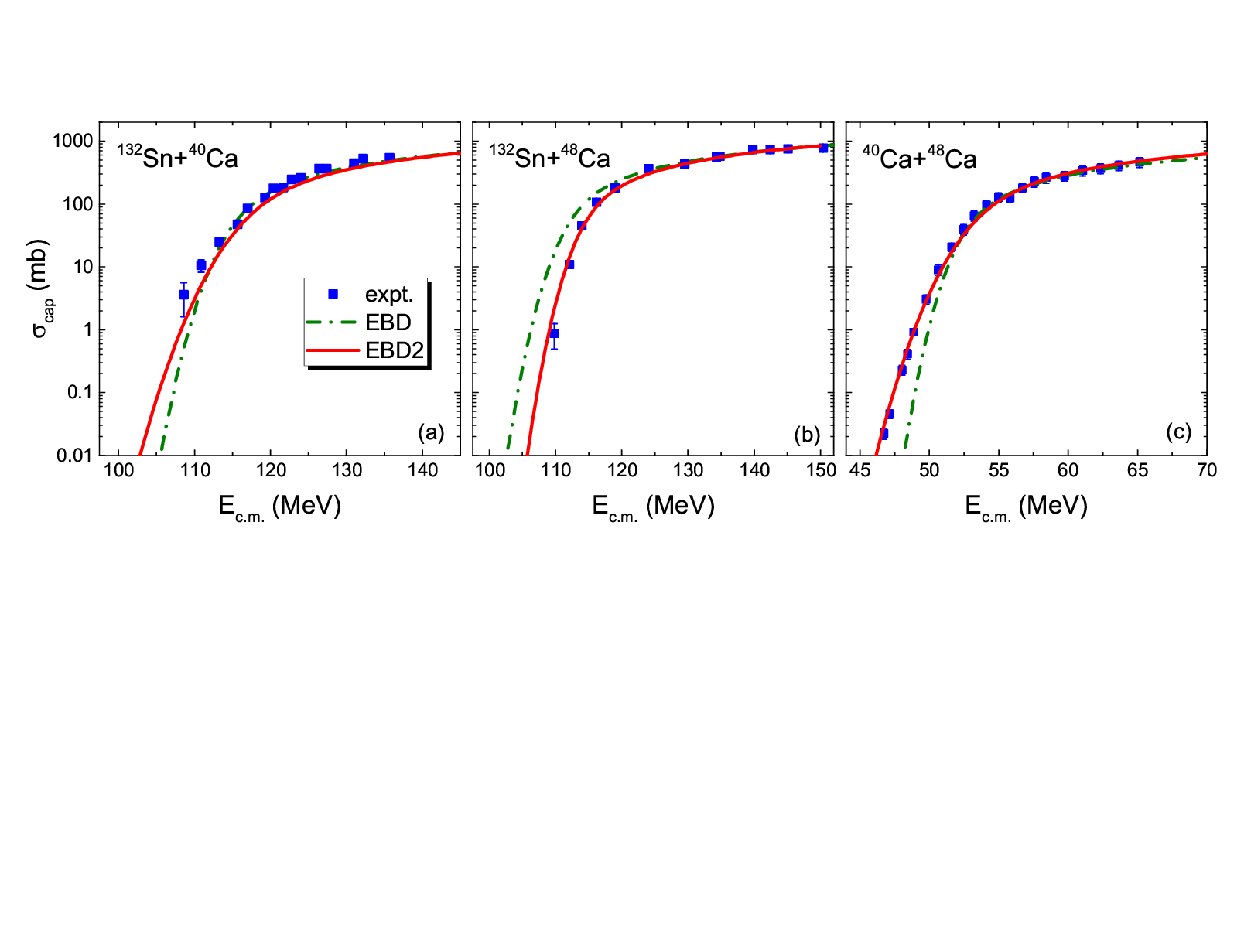}
	\caption{ (a) The same as Fig. 2, but for reactions $^{132}$Sn + $^{40,48}$Ca \cite{Kola12} and $^{40}$Ca + $^{48}$Ca \cite{Jiang10}. }
\end{figure}

To test the accuracy of EBD2 formula for describing the fusion reactions induced by neutron-rich doubly-magic nuclei, we show  the predicted capture (fusion) cross sections for  $^{132}$Sn+$^{40,48}$Ca and $^{40}$Ca + $^{48}$Ca in Fig. 3. Despite both $^{40}$Ca and $^{48}$Ca are doubly-magic nuclei, the $Q$-values in $^{132}$Sn+$^{40,48}$Ca are quite different (with $Q=-52.13$ MeV for $^{132}$Sn+$^{40}$Ca, and  $Q=-75.78$ MeV for $^{132}$Sn+$^{48}$Ca). Although with eight neutrons more in $^{48}$Ca, the excitation energy of the compound nucleus in $^{132}$Sn+$^{48}$Ca is only 37 MeV, smaller than that of $^{132}$Sn+$^{40}$Ca by about 26 MeV at $E_{\rm c.m.}=V_B$, due to the stronger shell effect in $^{48}$Ca and the isospin effect. In Ref. \cite{Yao24}, it is found that a relatively higher excitation energy at capture position can result in stronger effects for dynamical deformations and nucleon transfer in the capture process and broadens the width of the barrier distribution. In EBD2, the parameter $w_0$ is directly related to the excitation energy of the compound nuclei in fusion reaction. The obtained values of $W$ according to Eq.(5) are 3.83 MeV and 2.33 MeV for $^{132}$Sn+$^{40}$Ca and $^{132}$Sn+$^{48}$Ca, respectively. From the figure, one sees that the model accuracy is significantly improved in EBD2, indicating that the $Q$-value plays a role in the width of the barrier distribution. In addition, the correction term $\Delta B$ in Eq.(2) due to the competition between shell effect and isospin effect also plays a role for these reactions. For $^{132}$Sn+$^{48}$Ca, the barrier height is increased by $\Delta B=1.28$ MeV due to the competition between these two effects, with which the over-prediction of the cross sections from EBD at sub-barrier energies are obviously improved.

\begin{figure}
	\setlength{\abovecaptionskip}{ 0. cm}
	\includegraphics[angle=0,width=0.85\textwidth]{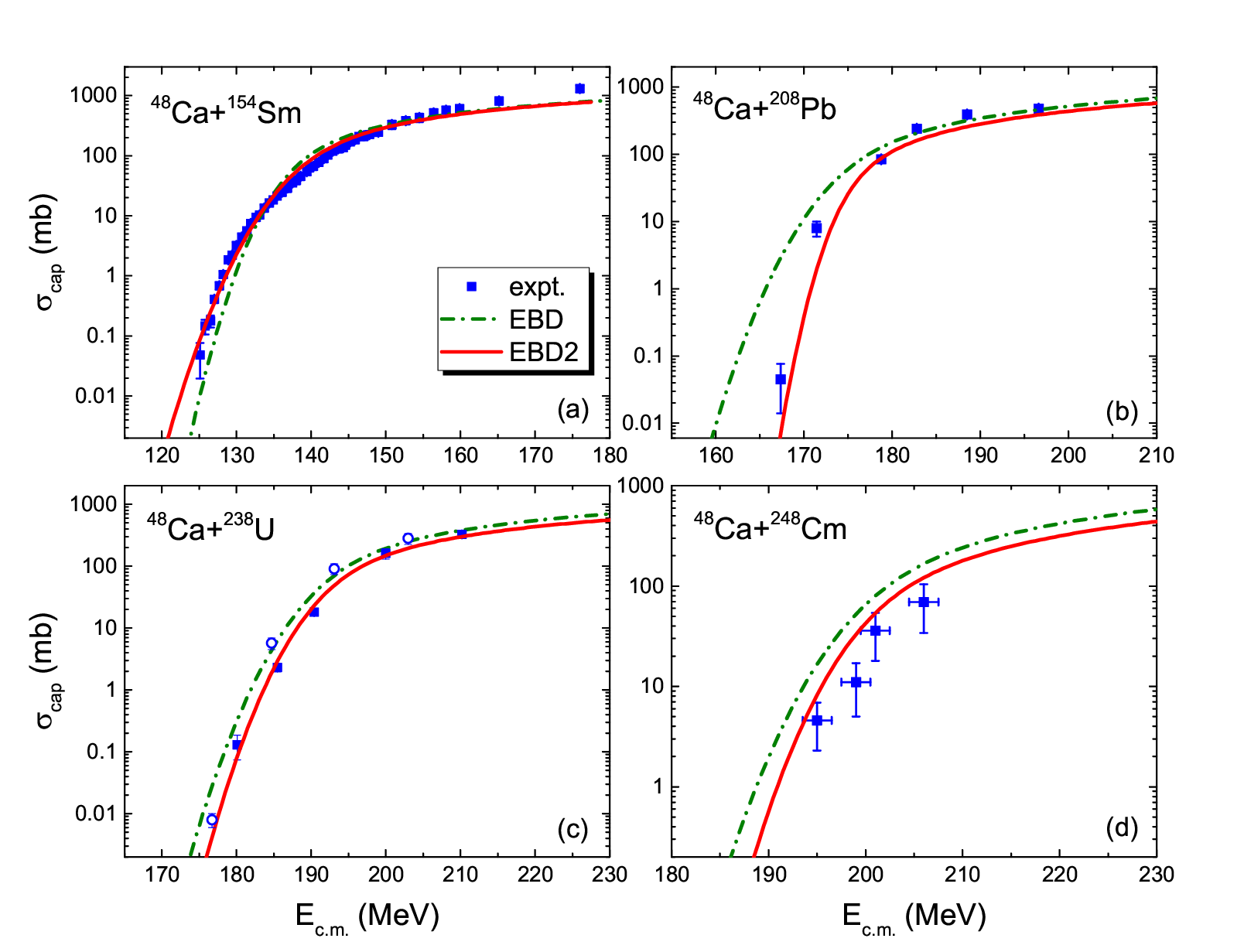}
	\caption{ (a) The same as Fig. 2, but for reactions  $^{48}$Ca+$^{154}$Sm \cite{Stef05},  $^{48}$Ca+$^{208}$Pb \cite{Prok08},  $^{48}$Ca+$^{238}$U \cite{Nishio12,Itkis22} and  $^{48}$Ca+$^{248}$Cm \cite{Itkis22}.}
\end{figure}

In Fig. 4, we show the predicted capture cross sections for four heavy fusion reactions induced by $^{48}$Ca. We note that the experimental data can be well reproduced by EBD2, particularly for $^{48}$Ca+$^{208}$Pb in which the experimental data are significantly over-predicted by EBD. The squares and circles in Fig. 4(c) denote the experimental data taken from \cite{Nishio12} based on position-sensitive multiwire proportional counters and \cite{Itkis22} based on two-arm time
of-flight spectrometer CORSET, respectively. For the reaction leading to the synthesis of super-heavy nuclei $^{48}$Ca+$^{248}$Cm, the extracted capture cross sections from the measured mass total kinetic energy distributions at energies around the barrier are evidently smaller than the predicted results from EBD. The results of EBD2 are better due to introducing the factor $F_{\rm DIS}$ to consider the influence of deep inelastic scattering on the barrier radius.

\begin{figure}
	\setlength{\abovecaptionskip}{ -2 cm}
	\includegraphics[angle=0,width=0.9\textwidth]{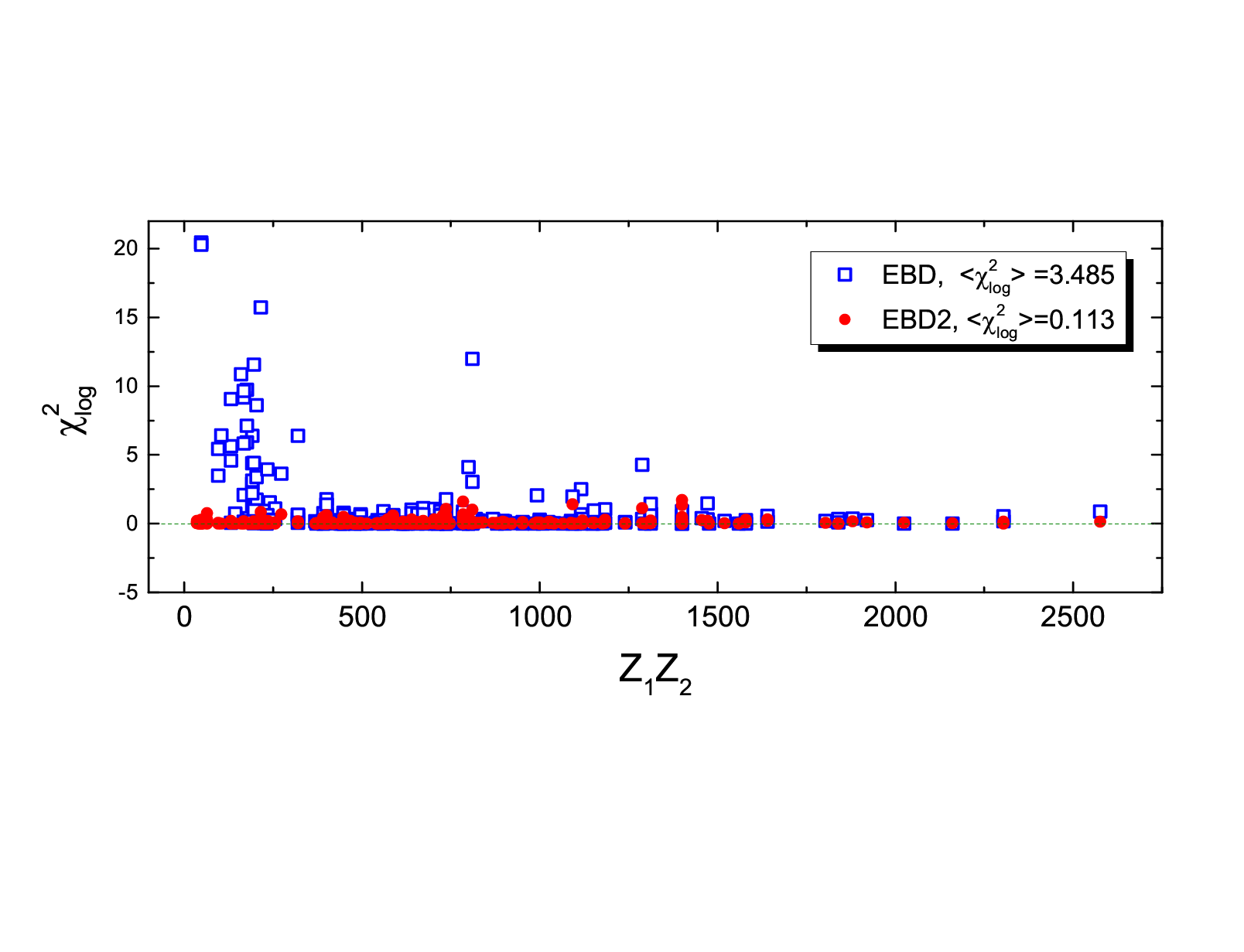}
	\caption{ Mean-square deviation between the predicted cross sections and the experimental data in logarithmic scale. The squares and the circles denote the results with EBD and EBD2, respectively.    }
\end{figure}

To further test the accuracy of EBD2, we systematically calculate the capture excitation functions for  426  fusion reactions with $ 35 < Z_1 Z_2 < 2600$. To analyze the model accuracy, we calculate the mean-square deviation between the predicted cross sections and the experimental data in logarithmic scale (which is called average deviation $\mathcal{D}$ in \cite{Wangbing}),
\begin{equation}\label{eq.x2log}
	\chi_{\log}^{2}=\frac{1}{N} \sum_{i=1}^{N}\left[\log \left(\sigma_{\mathrm{th}}\left(E_{i}\right)\right)-\log \left(\sigma_{\exp }\left(E_{i}\right)\right)\right]^{2}.
\end{equation}
Comparing with the traditional definition of $\chi^{2}$ in which the uncertainty of the cross section is involved, $\chi_{\log }^{2}$ is more effective to check the trend of cross sections at sub-barrier energies. In Fig. 5, we show the calculated mean-square deviation $\chi_{\log}^{2}$ as a function of the charge product $Z_1 Z_2$. One sees that the obtained mean-square deviations with EBD2 are much smaller than those of EBD, particularly for light fusion systems. For the total of 426 fusion reactions, the mean value of $\chi_{\log}^{2}$ is only 0.113 with EBD2, which is much smaller than the corresponding value with the original EBD (3.485).

To test the predictive power of the EBD2 formula, we simultaneously calculate the capture cross sections of two reactions \cite{Ban20} $^{12}$C+$^{248}$Cm and $^{16}$O+$^{244}$Pu  which are not included in the 426 sets fusion reaction data mentioned previously. We note that the measured capture excitation functions for these two reactions can be well reproduced by EBD2. In addition, we note that the fusion cross sections for some $\alpha$-induced reactions such as $\alpha$+$^{154}$Sm \cite{Gil85} at energies around the Coulomb barrier can also be described well by EBD2.

The capture cross section at a given center-of-mass energy $E_{\rm c.m.}$ can be also written as the sum of the cross section for each
partial wave $J$ \cite{Bao23},
 \begin{equation}
	\sigma_{\rm {cap} } (E_{\rm c.m.})=  \frac{\pi \hbar^{2}}{2 \mu E_{\rm c.m.}} \sum_{J} (2J+1)T(E_{\rm c.m.},J),
\end{equation}
$J$ represents the relative angular momentum and $T(E_{\rm c.m.},J)$ is the penetration probability of the two colliding nuclei overcoming the capture potential barrier in the entrance channel. The reduced mass of the entrance channel is simply given by $\mu= u A_1 A_2/(A_1+A_2)$. In the EBD2 formula, the corresponding penetration probability $T(E_{\rm c.m.},J)$ can be approximately expressed as \cite{EBD},
 \begin{equation}
	T (E_{\rm c.m.},J) \simeq  \frac{1}{2} \left[1+ {\rm erf} \left(\frac{E_{\rm c.m.}-B_{\rm eff} }{ \sqrt{2} \, W}\right) \right],
\end{equation}
with the effective barrier $B_{\rm eff}=V_B+ J(J+1)\hbar^2 /( 2 \mu R_B^2 )$. In Fig. 6, we show the predicted penetration probabilities and the corresponding capture cross sections for $^{54}$Cr+$^{243}$Am. From Fig. 6(a), one can see that the obtained penetration probabilities increase gradually from zero to one for head-on collisions and $T=0.5$ at the barrier energy 236 MeV, due to the introducing of the Gaussian barrier distribution. At $E_{\rm c.m.}=240$ MeV, the penetration probability decreases with the angular momentum and falls to 0.1 at $J=35$, which implies that super-heavy nuclei in this reaction are mainly formed at the central collisions. From Fig. 6(b), we note that the results with Eq.(7) and those with Eq.(1) are very close to each other, which indicates that the approximation in Eq.(8) is accurate enough. 

\begin{figure}
	\setlength{\abovecaptionskip}{ -2 cm}
	\includegraphics[angle=0,width=0.9\textwidth]{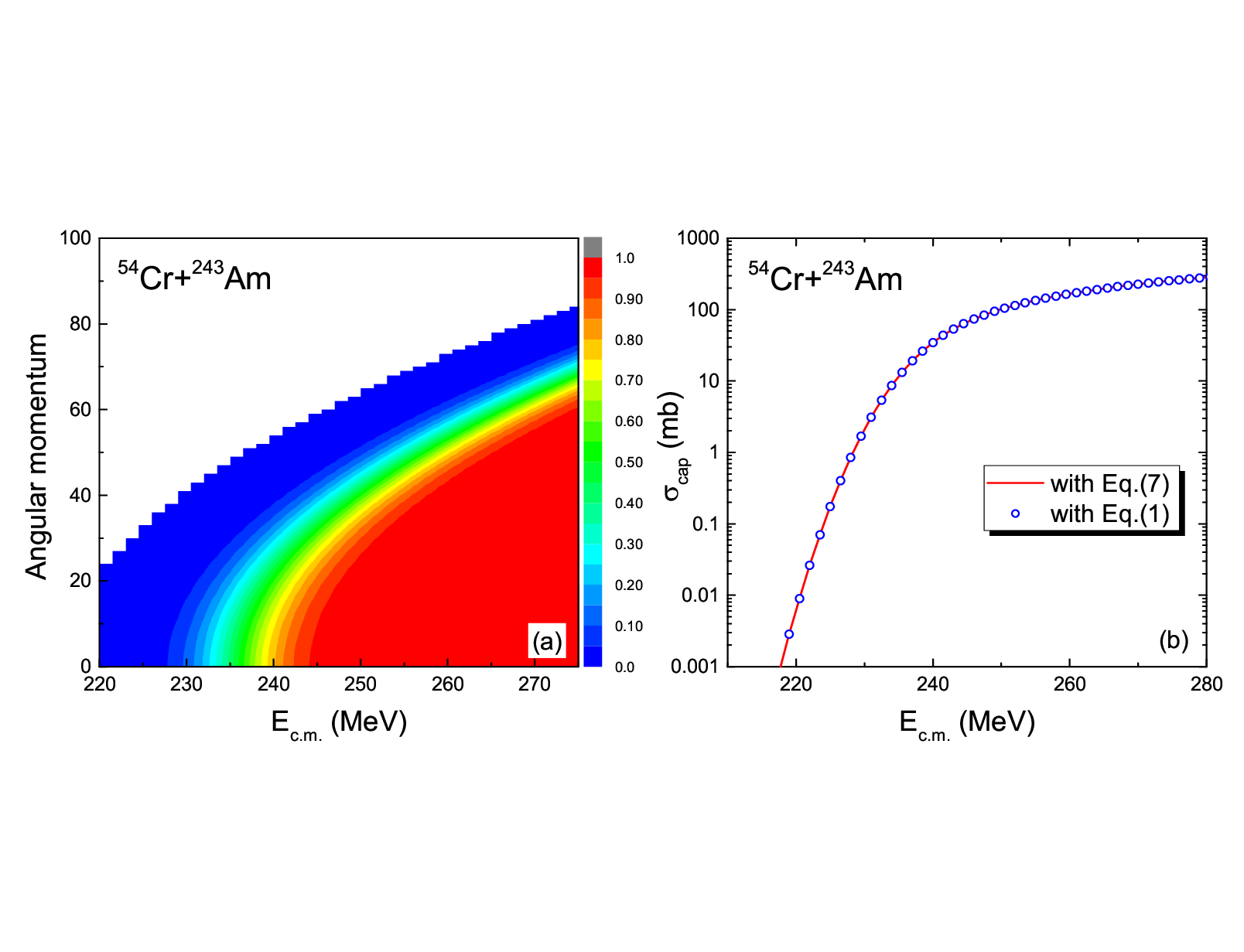}
	\caption{ (a) Penetration probability $	T (E_{\rm c.m.},J)$ with Eq.(8) for  $^{54}$Cr+$^{243}$Am. (b) Predicted capture cross sections for $^{54}$Cr+$^{243}$Am with the EBD2 formula. The solid curve and the circles denote the results with Eq.(7) and Eq.(1), respectively.    }
\end{figure} 

\begin{center}
	\textbf{IV. SUMMARY}
\end{center}

With a refinement to the input quantities of the empirical barrier distribution method, a much more accurate analytical capture cross section formula EBD2 is proposed. The key refinements involve (i) a surface correction to the parameter $z$ and a shell correction term for barrier height calculations, (ii) a $Q$-value-dependent width term to account for dynamical deformation effects, and (iii) a radius modification incorporating surface effects in light nuclei and deep inelastic scattering in superheavy systems. These refinements substantially improve the model accuracy. The average deviation (in logarithmic scale) between the predicted capture cross sections and the experimental data for a total of 426 reaction systems with $ 35 < Z_1 Z_2 < 2600$ is systematically calculated. EBD2 reduces the average deviation from 3.485 to 0.113. The surface correction term plays an important role in the fusion barriers for reactions between light nuclei, and a relatively higher excitation energy (related to the reaction $Q$-value) at capture position can result in stronger effects for dynamical deformations and nucleon transfer in the capture process, and thus broadens the width of the barrier distribution. The competition between shell effect and isospin effect is obviously for reactions with neutron-rich magic nuclei, such as $^{48}$Ca and $^{132}$Sn. For superheavy systems, the influence of deep inelastic scattering needs to be considered for a better description of the capture cross sections. Although the EBD2 method can systematically reproduce the data well, we still note that the measured cross sections for oxygen-induced reactions at above-barrier energies are slightly under-predicted by the EBD2 method, which could be due to the limitation of the single-Gaussian distribution in the calculations. By further adding a correction term from the classic cross section formula $0.1 \pi R_B^2(1-V_B/E_{\rm c.m.})$, the capture cross sections at above-barrier energies can be better reproduced. With the proposed EBD2 formula for better describing the capture cross sections, the evaporation residual cross sections for fusion reactions leading to the synthesis of superheavy nuclei could be further investigated with less uncertainties.

\begin{center}
	\textbf{ACKNOWLEDGEMENTS}
\end{center}
This work was supported by National Natural Science Foundation of
China (Nos. 12265006, U1867212). N. W. is grateful to Min Liu and Hui-Ming Jia for reading of the manuscript, and Jinming Chen for calculating the deviations $\chi_{\log}^{2}$. A capture cross section calculator based on the proposed EBD2 formula is available on http://www.imqmd.com/fusion/EBD2A.html

\end{document}